\documentclass[aps,superscriptaddress,twocolumn,a4paper,pra]{revtex4}
\usepackage{amsmath,amssymb,graphics,graphicx}

\begin{document}

\title{Entangling unitary gates on distant qubits with ancilla feedback}
\author{Kerem Halil Shah}
\affiliation{SUPA Department Of Physics, University of Strathclyde,
Glasgow G4 0NG, UK}
\email{k.halil-shah@strath.ac.uk}
\author{Daniel K. L. Oi}
\affiliation{SUPA Department Of Physics, University of Strathclyde,
Glasgow G4 0NG, UK}

\begin{abstract}
  By using an ancilla qubit as a mediator, two distant qubits can
  undergo a non-local entangling unitary operation. This is desirable
  for when attempting to scale up or distribute quantum computation by
  combining fixed static local sets of qubits with ballistic
  mediators. Using a model driven by measurements on the ancilla, it
  is possible to generate a maximally entangling CZ gate while only
  having access to a less entangling gate between the pair qubits and
  the ancilla. However this results in a stochastic process of
  generating control phase rotation gates where the expected time for
  success does not correlate with the entangling power of the
  connection gate. We explore how one can use feedback into the
  preparation and measurement parameters of the ancilla to speed up
  the expected time to generate a CZ gate between a pair of separated
  qubits and to leverage stronger coupling strengths for faster
  times. Surprisingly, by choosing an appropriate strategy, control of
  a binary discrete parameter achieves comparable speed up to full
  continuous control of all degrees of freedom of the ancilla.
\end{abstract}
\maketitle

\section{Introduction}

The ability to harness quantum phenomena for information processing
purposes underlies quantum computation (QC). There are several
different underlying computation models, including
gate-based\cite{GBQC}, measurement-based~\cite{RaussendorfBriegel},
adiabatic\cite{QuantumComputationByAdiabaticEvolution} and topological
models~\cite{KitaevAnyons}. Their interest in not only due to their
suitability to different physical substrates for implementation but
also on a more fundamental level as to the sets of resources necessary
or sufficient for universal QC.

Recently a subset of schemes have arisen based around the use of
ancilla systems, such as the ancilla-driven~\cite{ADQC,Twisted2012},
ancilla-control~\cite{ACQC} and quantum bus~\cite{QubusProposal}
proposals, where logical operations
are generated by interacting the qubits of the main register with an
ancilla system then performing operations on that ancilla system. The
various ancilla schemes are distinguished by their differing
requirements of the interaction between register and ancilla, the
operations on the ancilla and the number of required interactions. For
example, the ancilla-control scheme for implementing a single qubit
unitary on a register qubit requires being able to perform that
unitary on the ancilla~\cite{ACQC}; the ancilla-driven scheme requires
only arbitrary rotations about a single axis, provided the appropriate
measurement basis is available for measurements on the ancilla
\cite{ADQC}.

Ancilla driven quantum computation is particularly suited to the use
of hybrid physical
systems~\cite{QuantumRegisterNVCentre,UlmNVCentreAncilla} where there
is a memory register optimised for stability and long coherence times
and a short lived but easily manipulated ancilla system such as NV
centre nuclear-electron spins. The model of weaker or arbitrary
interaction strength is suited for when the interaction is not tunable
such as when dealing with scattering between flying and static qubits
\cite{FlyingQubits}. A stable memory register may be the product of a
particularly well chosen physical system but also it could be due to
the use of nodes of qubits that are egineered to perform
error-correction and fault tolerance codes locally as in several
proposals for networked quantum
computation~\cite{OiDevittHollenberg}. A distributed design may also
aid in parallelising circuit design for a time speed up or in aiding
scability of a physical
implementation.~\cite{RepeatUntilSuccessLinearOptics,DistributedQCCiracEkert}

In these cases different local nodes of qubits may have to be
connected by a different physical medium, such as a photon or coherent
beam system~\cite{DevittGreentreeHollenberg}. Therefore it is useful to
consider ancilla schemes in the context of distributed or networked
quantum computation where non-local operations are applied over
relatively large separations that inhibit coordination.

The problem of entangling a physically separated pair has been
considered before with methods such as the Barrett-Kok double
heralding approach~\cite{BarrettKokHeralding} where entangled states
are generated through projecting the system via photon pair
measurements or Lim, Barrett \emph{et al}
\cite{RUSQCUsingStationaryAndFlyingQubits}'s repeat-until-success
method through Bell basis measurements. In contrast, the operation on
the qubit pair in Ancilla Driven Quantum Computation remains a
reversible, commutable unitary gate that also requires only single
qubit measurements and does not require maximum entanglement with the
ancilla. This means that the process can be used in frameworks other
than the generation of cluster states for measurement based quantum
computation and can use non-maximal ancilla-register
interactions. However in the latter case, the process for generated an
entangling gate becomes stochastic.

This paper examines the use of feedback of ancilla measurement results
into subsequent generations. The application of control over the
ancilla state is also used to speed up this stochastic process and to make
it behave according to a well defined statistical behaviour. This
occurs in the broader theme of how we can trade off the requirement of
some resources at a cost of an increased time to implement specific
operations.

\section{Ancilla Driven Operations}

In an ancilla driven model, a qubit in a memory register interacts
with an ancilla qubit in a prepared state, then the ancilla is
measured and the resulting back-action on the register is unitary
depending on the parameters of the preparation, interaction and
measurement. If the interaction is locally equivalent to
$e^{-i\frac{\pi}{4} \sigma_z\otimes\sigma_z}$ (CZ type) or
$e^{-i\frac{\pi}{4}(\sigma_x\otimes\sigma_x+\sigma_y\otimes\sigma_y)}$
(CZ.SWAP type) then one can generate an arbitrary rotation angle
$\beta$ on the register qubit by performing a rotation by $\beta$ n
the ancilla before measurement. E.g. Using a CZ gate and an ancilla
prepared in the $|+\rangle$ state, performing a rotation about the
$\hat{x}$ axis, $R_{\hat{x}}(\beta)$, on the ancilla then measuring in
the 0/1 basis; this enacts $Z^jR_{\hat{z}}(\beta)$, where $j=0,1$ is the
measurement result, on a single register qubit.

Rotations about a single axis will of course not be able to generate
any arbitrary single qubit unitary. However if the interaction is
$(H_a\otimes H_r).\text{CZ}$, where we have included Hadamard local
gates as part of the interaction, then by accounting for the extra
local effects on the ancilla by applying $J(\beta)=H.e^{-i\beta
  \sigma_z}$ instead before the measurement, $X^j.J(\beta)$ acts on
the register. The class $J(\beta)$ allows one to feed forward the
measurement results to commute through the Pauli correction into a
single post-correction and to apply any single qubit unitary (up to
global phase), $U\equiv J(0)J(\alpha)J(\beta)J(\gamma)$ for some
$\alpha,\beta,\gamma$~\cite{ParsimoniousAndRobust}.

\begin{figure}
\includegraphics[width=\columnwidth]{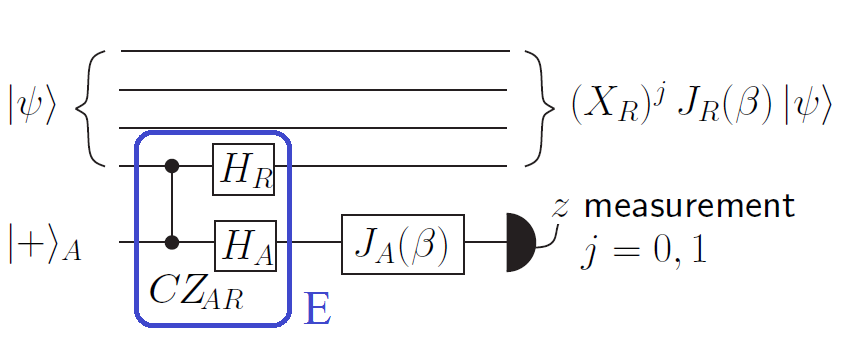}
\caption{An circuit diagram example of ancilla driven quantum
  computation using the connection interaction $(H_A\otimes
  H_R).CZ$~\cite{ADQC}. The ancilla is prepared in a fixed state, the
  choice of measurement basis is represented by the application of the
  unitary gate $J(\beta)$ to the ancilla in order to draw attention to
  the symmetry between the action on the ancilla and the register.}
\label{fig:ADQCpaperSingleQubit}
\end{figure}

Crucially, by applying this same interaction between the ancilla and
two subsequent register qubits, a CZ gate can be generated on the
register, up to local gate corrections, thus providing the resources
for universal quantum computation.

\begin{figure}[h]
\includegraphics[width=\columnwidth]{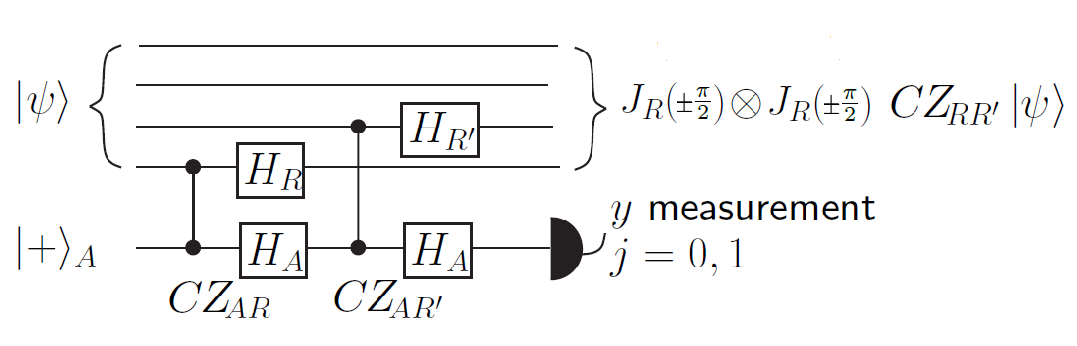}
\caption{A two qubit gate in the ADQC scheme, using the same
  connection interaction twice, enacts a deterministic entangling gate
  with probabilistic local unitary effects. Here, the choice of
  measurement basis is fixed.}
\label{fig:ADQCpaperTwoQubit}
\end{figure}

However if the interaction is instead equivalent to $e^{-i\alpha
  \sigma_z\otimes\sigma_z}$ or
$e^{-i(\alpha_x\sigma_x\otimes\sigma_x+\alpha_y\sigma_y\otimes\sigma_y)}$
for $0<\alpha<\frac{\pi}{4}$ then the rotation angle $\beta$ becomes
random in a way without simple post-corrections~\cite{KHSOi01}. This
applies also for the two qubit entangling gate: a gate locally
equivalent to a Control $\gamma$ rotation, $C(\gamma)$, is generated
with random $\gamma$ depending on the measurement result.

\begin{figure}[h]
\includegraphics[width=\columnwidth]{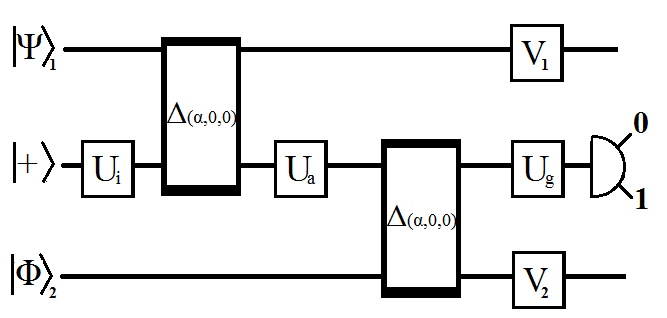}
\caption{A circuit for generating a two qubit control-unitary gate
  with an interaction parametrised by an arbitrary coupling strength
  $\alpha$: $\Delta_\alpha=e^{-i\alpha
    \sigma_z\otimes\sigma_z}$. Local gate differences on the ancilla
  can be accounted for by the setting of $U_i$, $U_a$ and $U_g$. Local
  gate effects on the register can by accounted by a measurement
  dependent post-correction $V_{1/2}$. $V_{1/2}$ will commute with
  $\Delta_\alpha$ thus several applications can be treated with a
  single post-correction.}
\label{fig:TwoQubitEntanglementCircuit}
\end{figure}
 
It may however be possible to generate a chosen $C(\gamma)$ gate if a
probabilistic achievement time is allowed. Any gate generated by the
use of a connection interaction in the local equivalence class of
$e^{-i\alpha \sigma_z\otimes\sigma_z}$ will also be able to be
diagonalised in the computational basis by local unitary gate
operations. If the local operations can be directly created,
or created by the use of well engineered qubits and interactions in a
local node, then the diagonalised products of each generation will all
commute and the random behaviour will map to a random walk on a circle.
The gates
generated would be of the general form
$diag(e^{i\phi_1},e^{i\phi_2},e^{i\phi_3},e^{i\phi_4})$ which is
locally equivalent to a Control-$R_{\hat{z}}(\Phi)$ gate where
$\Phi=\phi_4-\phi_3-\phi_2+\phi_1$. The angles are mapped to a point
on a circle. So at each time interval, an angle is randomly gated to
represent the gate generation and is added to the sum of all previous
gate generations; if the new sum lies with a target region, given by
$\pi \pm \epsilon)$, for some chosen error $\epsilon$, the operations
are halted.

\section{Unguided behaviour of random generation}

The probabilistic nature of the generation of gates in this scheme
leaves us with the problem of understanding the statistics of the time
it takes to reach any arbitrary gate.

\begin{figure}
\includegraphics[height=0.645\columnwidth,width=\columnwidth]{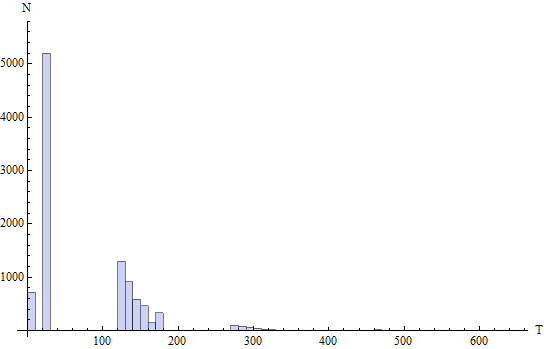}
\caption{The distribution of the target region hitting times given by
  the simulation of the gate generation with an interaction with
  $\alpha=\frac{\pi}{16}$. The mean hitting time is 74.1, the standard
  deviation is 74.5.}
\label{fig:HittingTimes}
\end{figure}

We simulated the creation of a CZ equivalent gate by use of the
circuit in figure \ref{fig:TwoQubitEntanglementCircuit} with the
choice of $U_i$=$\mathbb{I}$, $U_a=R_{\hat{x}}(\frac{\pi}{2})$ and
$U_g=H$ i.e. preparation and measurement in the $X$ eigenstate
basis. These settings are not unique to the gates generated. This was
performed 10,000 times, with an error bound of $\pi/100$ and the
resulting distribution is displayed in figure \ref{fig:HittingTimes}.

The distribution of hitting times, when put into sufficiently broad
bins, can be described by an exponential tail. One of the results is
an irregular angle that depends on the coupling strength while the
other is $\pi$ for all couplings. Aside from the initial probability
of success in the first step by generating the angle $\pi$, we expect
that one of the results could be used to reach the target within the
bound if applied a large number of times scaling with the error
according to the dimensions of the space, in this case,
linearly. Indeed, we found a linear scaling.

A result would be expected at around the time where this large number
matches of the expected number of successes of that gate and the
number of $\pi$ results generated is even. The probabilistic nature of
this causes a distribution around this point so there is a length of
time associated with an opportunity for success. This then repeats in
a decaying tails. This probability of success in a fixed time interval
ends up accruing the properties of a similarly described probability
distribution, the geometric distribution. This is not an exact
description but we will later find that it allows a comparison to a
case with an exact solution.

When employing a strategy for feedback, it is necessary to have a
picture in which the control parameters relate to the ultimate
property desired for optimisation: the number of gate generations
required. Due to the decaying tail distribution of hitting times in
the case without feedback, we expect that an important feature of the
hitting time statistic is the minimum number of steps required to
create a finite probability of hitting the target. In the following
section, we discuss strategies based around the principle of
optimising the probability of success in a minimal number of steps.
Because there is always a probability of generating $CZ$ with any
coupling by preparation and measurement in the $X$ eigenstate basis,
it is possible to perform a ``one step" strategy: maximise the
probability to hit the target in the next step.

\section{Strategies for guided gate generation}

In the ancilla-driven scheme, an ancilla is prepared in a specific
basis, undergoes an interaction with one register qubit using a
specific connection interaction, the ancilla is then interacted with
the second register qubit and then measured in an appropriate
basis. The ancilla is controlled by the choice of preparation state
and measurement basis provided that any local unitary actions on the
ancilla in the intermediate time between interactions account for the
conditions for a unitary, entangling generated gate on the
register. This restriction results in only two degrees of freedom. In
the context of a spatially separated pair in a distributed network,
the two parameters can be seen as one requiring the preparation of the
ancilla performed by Alice and the other the measurement performed by
Bob. If you have only 1 degree of freedom then the task of setting up
the strategy can be placed on only one partner. Alice could prepare a
sequence of qubits and then send them to Bob. Bob then only has to
measure them in a fixed basis with minimal instructions about what to
do after a certain measurement result. We also consider how a strategy
might affect the complexity of Bob's instructions. Since Bob only has
two measurement results, he must at some point receive a string of
binary with the instruction to stop at the point when the instructions
don't match the string; perhaps those so interested could examine all
possible strings and the entropy of instructions for particular
strategies, we will just be focusing on a question surrounding a
simple case where the stopping condition is always the same
measurement result for Bob. If Bob was looking for signals coming out
of two ports, Bob would just wait until he one of those ports thus we
call this a "one port" strategy. This might also be important in an
experimental context if the measurement process is prone to a
particular measurement bias.

So we consider
\begin{itemize}
\item What if we have control over only 1 degree of freedom instead of 2?
\item What if we use only 1 port instead of 2?
\end{itemize}
Since the process is probabilistic, we are interested not just in the
expected number of ancilla one side may have to send to another but
more the total number that have to be prepared to ensure a certain
probability of success. This also can be seen as a division of tasks-
Alice prepares a number of ancilla qubits $N \text{s.t.} P(n<N)\geq
0.999$ and transmits them to Bob who then has to carry out the
instructions for when to stop permitting the ancilla to interact with
his qubit. Based on approximately geometric behaviour, the expectation
time is linearly inverse to the probability per time interval and
naturally linear with the time interval. So if the time interval is
increased by $n$ steps, the probability needs to be increased by
$n$. The value of $N$ can also be approximated by a linear multiple of
the expectation value so they enforce a restriction on when a ``multi
step" strategy is viable- a two step strategy should double the
probabilities and so on.

\subsection{The one step strategy}
At each step set the conditions so that one measurement result
generates a gate which corresponds to the angle difference between the
present point on the circle and the point $\pi$. If this measurement
does not occur, find the distance between the present point and the
point $\pi$ and attempt to generate that gate. Upon every failure,
find the new distance between the target and the current point and
attempt to generate that gate.
\begin{figure}[h]
\includegraphics[height=0.71\columnwidth,width=0.7\columnwidth]{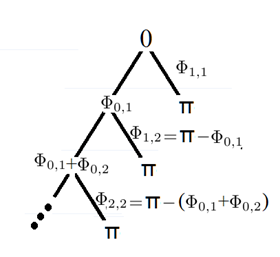}
\caption{The probability tree of the ``one step" strategy. The first
  step will always require generating a CZ equivalent gate, the other
  result will be dependent on the coupling and will then dictate all
  future conditions. The conditions are reset at each step with each
  new ancilla.}
\label{fig:ProbTree}
\end{figure}

\begin{figure}
\includegraphics[height=\columnwidth,width=0.79\columnwidth]{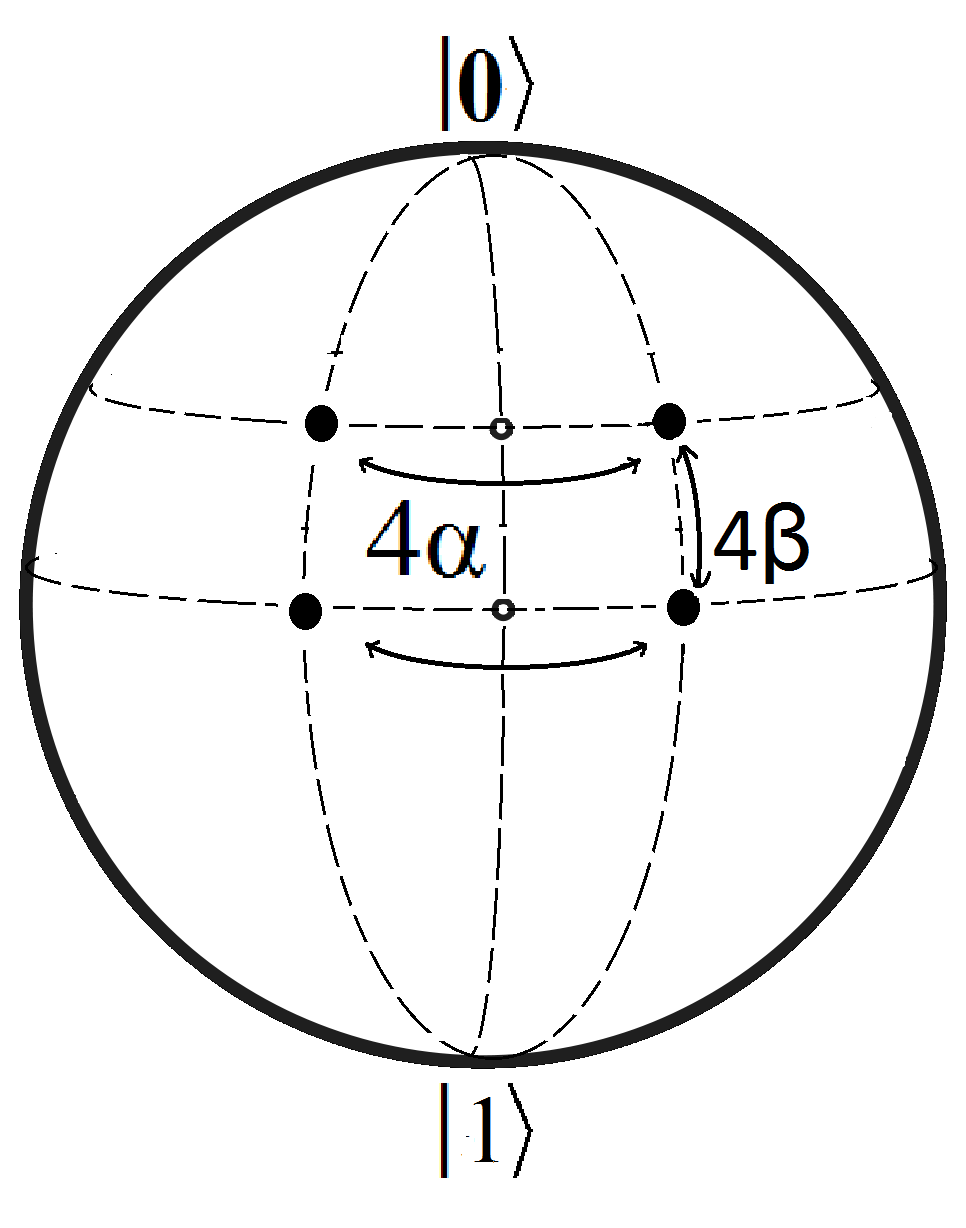}
\caption{The final state of the ancilla can be seen as a mixture of
  four points on the Bloch sphere. In order for the post-measurement
  action to be unitary the four points must lie on a circle and the
  measurement basis axis go through the centre. The cap sizes dictate
  the probability of the results, the distribution of the points
  around the circle affects the entangling power of the gate
  generated. The result within the minor cap is the more likely but
  will generate a less entangling gate. By adapting the parameters
  $\beta$ and $\theta$ these two aspects can be controlled.}
\label{fig:BlochSpherePicture}
\end{figure}

Understanding the setting of the conditions of the gate generation can
be understood with a minor review of the Bloch sphere picture of the
entanglement condition (see figure
\ref{fig:BlochSpherePicture}). Since the non-local part of the Cartan
decomposition of the connection interaction is diagonal in the
computational basis, the ancilla before measurement has evolved as
$|a\rangle \sum_{ij} c_{ij}|i\rangle|j\rangle \rightarrow
\sum_{ij}c_{ij}^\prime |a_{ij}\rangle|i\rangle|j\rangle$. The final
states $|a_{ij}\rangle$ will be
\begin{equation}
\text{cos}\left(\frac{\theta-(-1)^i 2\beta}{2}\right)|0\rangle + e^{i(-1)^j 2\alpha}\text{sin}\left(\frac{\theta-(-1)^i 2\beta}{2}\right)
\end{equation}
These will map to four points on the Bloch sphere. The angle $\beta$
must be set by operations before and after the first interaction, the
angle $\theta$ is set before the second interaction and must be known
so that a measurement can be applied which is mutually unbiased to all
four points.

Given any coupling strength, at the start of the strategy, the first
attempt to generate $CZ$ is performed the same way: the ancilla is
prepared in the $+$ state and then measured in the ${|\pm\rangle}$
basis with the ``$-$" (port 1) result generating a gate equivalent to
$CZ$ ($C(\pi)$). The ``$+$" (port 0) result would generate a blow back
gate $C(\Phi_0)$. There is a sense of direction with the gate
generation; one port gives $C(-|\Phi_0|)$, the other $C(+|\Phi_1|)$,
clockwise or anticlockwise around the circle that represents the
$C(\gamma)$ group. One can simply switch the direction association of
the ports by performing a bit flip either immediately before or after
transmission from Alice to Bob, so we will ignore the exact sign
requirements in the notation from here on and simply note the need to
flip. Having travelled ``clockwise", the best next step is to continue
in that direction and generate $C(\pi-|\Phi_0|)$. If $\Phi_0$ is small
then $\pi-\Phi_0$ will be large enough that it can only be generated
from port 1, the port with larger $\Phi$ but smaller probabilities
upper-bounded by $\frac{1}{2}$. Another feature of port 1 is that the
probability increases as the preparation and measurement variables
($\beta,\theta$) are increased and for a fixed $\Phi_1$, $\theta$
increases with $\beta$. Therefore the for optimal probability, it is
only needed to fix one of these parameters to the maximum and vary the
other. So an 1 port strategy is effectively also a 1 degree of freedom
strategy where the only task is finding the gate and the parameters
for the next step.

At every step $n$, there is one gate that matches success
$C(\Phi_{1,n})$ and a failure gate $C(\Phi_{0,n}$, therefore to be at
step $n$, the current action on the register system is the product of
previous failures $C(-|\Phi_{0,1}|+\Phi_{0,2}+...+\Phi_{0,n-1})$. The
next gate to be generated for success must be
$C(\pi-(|\Phi_{0,1}|-\Phi_{0,2}-...-\Phi_{0,n-1})$.

The magnitude of the angle $\Phi$ of both ports increases with the
probability of success of $\Phi_1$. So because the largest angle to be
generated is $\pi$ in the first step, the first step has the highest
probability of success and also the highest value of the failure gate
$\Phi_0$. Therefore $\pi-|\Phi_{0,1}|$ is the smallest value and has
the smallest probability of success. These two first values provide a
bound on the behaviour of the strategy. The cumulative distribution
function based measure, $P(n<N)$, can be compared to the CDF of
constant probability for each step using the extreme probabilities:
$1-(1-p_2)^n<P(n<N)<1-(1-p_1)^n$.

This also means that the first step provides the threshold for when a
two port strategy is viable: when is $\pi-|\Phi_{0,1}|$ small enough
that it can be generated from port 0? Since $\Phi_{0,1}$ is also the
maximum $\Phi_0$, it must be when $\Phi_{0,1}=\frac{\pi}{2}$. Port 0
has a different $(\beta,\theta)$ for fixed $\Phi_0$ relationship and
it's probabilities are optimised away from the fixed measurement
conditions so this is also the threshold for when a 2 degree of
freedom strategy can be involved.

\subsection{The ``flip-undo" strategy}
Up until now we have discussed the ability to manipulate the ancilla
with the assumption that we can exercise any arbitrary single qubit
unitary gate. This is in line with the requirements of the
ancilla-driven and ancilla-control schemes. However we have also
developed a scheme for exploring what can be done with as simple an
action on the ancilla as possible: we have only available to us a
fixed preparation state, a fixed measurement basis and the choice of
whether or not to implement a bit flip gate, $X$- specifically the bit
flip required to change the sense of direction of the ports. What this
provides is the ability to attempt to undo a previous action hence the
designation the ``flip-undo" strategy.

After attempting to generate a $C(\pi)$ gate in a single step, if the
result failed, attempt to go back to the origin. Whether you have
arrived at the origin or not, attempt to generate a $C(\pi)$ gate with
the product of the next generation. If one fails again, repeat the
process from the second step. Repeat until success.
\begin{figure}[h]
\includegraphics[height=0.81\columnwidth,width=0.7\columnwidth]{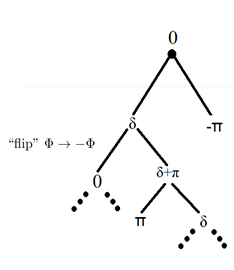}
\caption{The probability tree of the ``flip-undo" strategy receives
  all possible points in the strategy after 2 steps. After the first
  step, it can be seen as a repeat-until-success strategy where the
  time to repeat is 2 gate generations.}
\label{fig:FlipUndoTree}
\end{figure}

The inspiration for this scheme comes from questioning why the two
qubit gate is equivalent to $C(\gamma)$ and not $C(-\gamma)$. The
answer comes from the Bloch sphere picture of the four possible states
of the ancilla after interaction and their orientation. In the middle
of the procedure, it is only two states dependent on the first
register qubit: $\sum_{i}|a_i\rangle |i\rangle \sum_{j}
c_{ij}^{\prime\prime} |j\rangle$. On the Bloch sphere, these will be
two points, one above the other on the same vertical plane, in order
for the conditions for the gate to be unitary and entangling to be
fulfilled. Which is above which determines the sign of the rotation
angle. So if one was able to flip the orientation the sign would
change. This can be done by introducing an $X$ gate on the ancilla in
between the two connection interactions.

It should seem obvious that in a case where either $C(\pi)$ or
$C(\gamma)$ is generated and the target is $CZ=C(\pi)$ that it is
preferable to label a result $C(\gamma)$ a failure and attempt to undo
it in order to try again to generate $CZ$ directly. Yet that then
creates a possible result where the sequence product is
$C^\dagger(\pi)C(\gamma)=C(\gamma+\pi)$. Again the apparent best
decision is to attempt to undo $C(\gamma)$ as this will now
immediately lead to the target gate. In the following step, the only
two possible product sequences must result in $C(\pi)$ or $(C(\gamma)$
which makes employing this strategy form a closed loop.

Now that there is a description and probability tree for a finite
number of points on the circle, we can find an exact description of
the time statistics using the recursive relationships between the
expectation times at different points, if we take the probability of
generating $C(\pi)$ in the first step to be $p$:

\begin{align*}
\bar{n}=& p+(1-p)(\bar{n}_1+1) \\
\bar{n}_1=&p(\bar{n}_2+1)+(1-p)(\bar{n}+1) \\
\bar{n}_2=&p(\bar{n}+1)+(1-p) \\
\Rightarrow& \bar{n}=1+\frac{1}{p}
\end{align*}

Another way to look at it is that after the probability of success in
the first step, there is a 2 step time interval which results in a
probability of success of $2p(1-p)$ which can be repeated. So an
\emph{exact} geometric distribution tail is formed where the expected
time is $2.\frac{1}{2p(1-p)}$ and so
$\bar{n}=p.1+(1-p).(\frac{1}{p(1-p)}+1)=1+\frac{1}{p}$. The cumulative
density function is given by $1-(1-p)(1-2p(1-p)^k$ where the number of
transmitted ancilla qubits is $2k+1$.

\section{Numerical results}
We found the parameters and resulting probabilities for continuing
with the one-step strategy for 500 steps for a range of coupling
strengths of the connection interaction. The full range for
$0<\alpha<\frac{\pi}{4}$ was covered for the 1 port 1 degree of
freedom strategy where the degree of freedom was represented by the
preparation parameter $\beta$. We then found more values for the range
of coupling strengths that starts just before the threshold for the
two port strategy. In this range we then found the values for a two
port strategy where one could only vary $\beta$ and perform no
optimisation of port 0 and then found them again for when optimisation
over $\beta$ \& $\theta$ is allowed. Finally we checked for just the
one port strategy, the probabilities for each step when the
measurement parameter is the allowed of degree of freedom rather than
the preparation and this did turn out to give the exact same results.

\begin{figure}[h]
\includegraphics[height=0.60\columnwidth,width=\columnwidth]{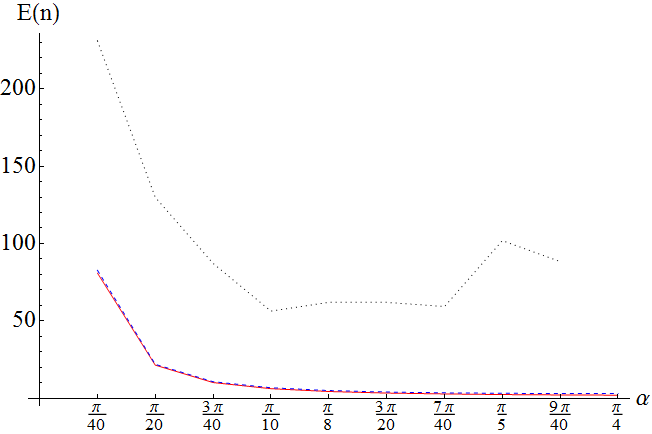}
\caption{A comparison of the unguided (dotted black) generation
  against the one step (solid red) and flip-undo (dashed blue)
  strategy with their expected hitting times against coupling
  strength. Relative to the unguided approach, the flip-undo strategy
  is nearly as completely effective as the one step strategy while
  required less ancilla control. The unguided expectation times are
  not well correlated with the coupling strength: while significant
  differences in step size from large coupling differences do impact
  on the number of steps, the ability to approach arbitrarily close to
  the target relates to the difference of the step size with rational
  divisions of $\pi$ making for small scale chaotic behaviour.}
\label{fig:ExpectationTimes}
\end{figure}

\begin{figure}[h]
\includegraphics[height=0.60\columnwidth,width=\columnwidth]{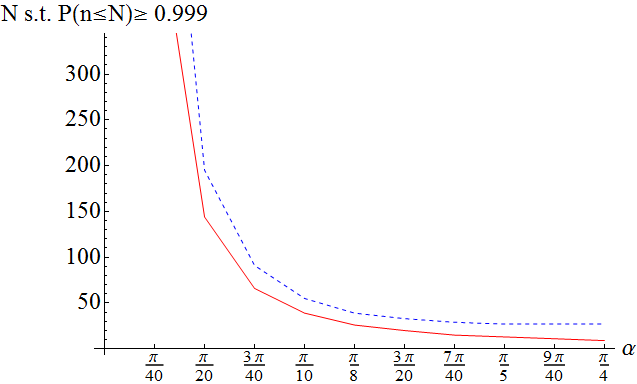}
\caption{A comparison of the one step (solid red) strategy against the
  flip-undo (dashed blue) strategy: the number of ancilla that need to
  be prepared to guarantee a 99.9\% chance of success.}
\label{fig:FlipUndoPlot}
\end{figure}

\begin{figure}[h]
\includegraphics[height=0.6\columnwidth,width=\columnwidth]{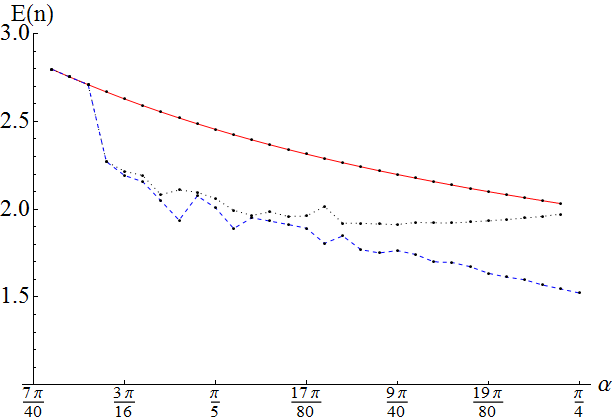}
\caption{A comparison of strategies for different numbers of degrees
  of freedom at high coupling strengths. The solid red curve
  represents the 1 port/1 d.o.f. approach, the dotted black curve is
  the 2 port/ 1 d.o.f. strategy and the dashed blue curve is the 2
  port/2 d.o.f. strategy. The threshold occurs at approximately
  $0.73\frac{\pi}{4}$.}
\label{fig:PortStrategyComparison}
\end{figure}

The order of improvement between different one-step strategies is less
than a single step in the expectation time. The 1 port strategy tends
towards an expectation number of 2 while the 2 port/ 2 degree of
freedom strategy tends toward 1.5 ; the 2 port/ 1 d.o.f. approach has
a peak in improvement near the middle of the range but at very close
to the maximum coupling returns back to the 1 port strategy. This
scale of improvement can be expected from the behaviour of the
probability of either port. As $\alpha \rightarrow \frac{\pi}{4}$,
$\Phi_{0,1}\rightarrow -\pi$, $\pi-|\Phi_{0,1}|$ becomes very small
and thus the probability out of port 0 in the second step tends to
1. Any consideration of multi-step strategies in this range is
therefore of limited advantage- the behaviour where failure in the
first step improves the probability of success in the second step
which we would expect to be a feature of any two step strategy is
already a feature of the one step strategy with two ports and access
to both parameters.

\begin{figure}
\includegraphics[height=0.65\columnwidth,width=\columnwidth]{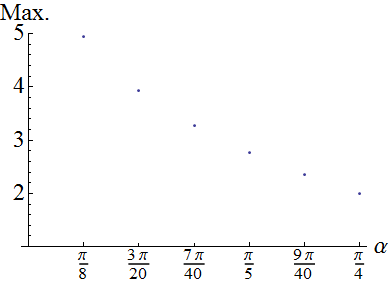}
\caption{The maximum number of steps in a strategy as allowed by the
  hard limit of $\frac{1}{p}$ for a given coupling, displayed over the
  top half of the range of coupling strengths.}
\label{fig:OneOverP}
\end{figure}

The viability of a multi-step strategy at the lower coupling step
range can be examined using the lower bound on the probabilities of
success in each step found from the probability in the second
step. This value describes the behaviour of a geometric distribution
that bounds the behaviour of the one step strategy; for a multi-step
strategy to be effective it must at least improve upon this and since
a multi-step strategy takes place over $n$ steps, it must improve the
probability by at least a factor of $n$ yet this will be limited by
the maximum value of 1. In figure \ref{fig:OneOverP} we can see what
the maximum possible number of steps for a multi-step strategy could
be for any improvement to be possible.

The most striking result is that the ``flip-undo" strategy has very
little cost in the expectation time compared to the one-step
strategy. The gap between it and the 1 port approach only approaches a
maximum of one step however the effect is more significant when
considering the minimum number of ancilla required to secure $P(n\leq
N)\geq 0.999$ but the relative effect is diluted as the coupling
strength gets weaker.

\section{Conclusion}

In summary, we have analysed a implementation of a maximally
entangling gate between distant qubits mediated by interaction with
flying ancilla with an arbitrary but fixed coupling strength. Due to
the stochastic nature of the measurements and the non-determinism of
the induced gate sequence, the time required for success is
random~\cite{KHSOi01}. By use of feedback on the ancilla preparation
or the measurement basis, some improvements can be made over a
stationary random walk strategy.


We have examined how the addition of local unitary gate control on the
ancilla qubit can speed up the expected time for implementation and
reduce the total number of ancilla qubit required for a given
fidelity. What has been found is that the improvement from control
over additional degrees of freedom is small which may be important in
the context of distributed or networked quantum computation.

If the task is distributed between two separated devices,
co-ordination between the devices only allows for some speed up past a
threshold coupling strength. The eventual speed is small and indeed
the benefits of applying any control and feedback can be mostly
realised by the inclusion of only one single extra operation on the
ancilla: the ability to choose to apply a bit flip. The dominant
factor appears to be the group structure of the gates that are
generated during the process and the ability to apply a bit flip to
the ancilla ensures that only four possible gates can be generated
which leads to a speed up over the generation of a continuous group.

Yet to be investigated are using interactions of the class
$e^{-i(\alpha_x\sigma_x\otimes\sigma_x+\alpha_y\sigma_y\otimes\sigma_y)}$
to generate gates in its own class. However the abelian structure and
single parameter of the $C(\gamma)$ gates has been a large part of the
simplification of the analysis and possible speed up and one would
expect that by having a two-parameter target, one would square the order
of the expectation times. A two parameter interaction would be better
created by applying two single parameter interactions with local
unitary gates between them.

The strategies employed seek to minimise the communication between two
parties attempting to generate a shared entangling gate. The strategy
can be worked out by A and instructions transmitted to B before
sending any ancilla; the local post-corrections can be commuted
through so B's instructions can be transmitted after all have been
sent. This allows us to envisage a scenario in which A sends B a
packet of $N$ ancillae where $P(n<N) \geq 0.999$. This avoids latency
in classically transmitting results and instruction between
ancillae. This does however require that any potential implementation
allows for B to be able to feed the ongoing measurement results into a
pre-interaction local operation in some strategies. Once B has hit the
target gate, B would have to be able to prevent further interactions
(whether by turning the interactions off or applying an appropriate
pre-correction) for the rest of the ancillae in the transmitted
packet. Thus times allowing for local corrections will limit the
transmission rate. Considering the minimal speed up from doing
otherwise, this speaks to the advantage of keeping to a one degree of
freedom strategy.

\bibliographystyle{unsrt}
\bibliography{bibliography}

\end{document}